\documentclass[aps,a4,twocolumn]{revtex4}
\pdfoutput=1
\usepackage{amsmath}
\usepackage{latexsym}
\usepackage{float}
\usepackage{amssymb}
\usepackage{graphicx}
\usepackage{epsfig}

\begin{document}

\title{New Development of Monte Carlo Techniques for
Studying Bottle-brush Polymers}

\author{Hsiao-Ping Hsu}

\affiliation{Institut f\"ur Physik, Johannes Gutenberg-Universit{\"a}t Mainz 
Staudinger Weg 7, 55099 Mainz, Germany}

\begin{abstract}
Due to the complex characteristics of bottle-brush polymers, it became a 
challenge to develop an efficient algorithm
for studying such macromolecules under various solvent conditions
or some constraints in the space by using computer simulations.
In the limit of a bottle-brush polymer with a rather stiff backbone 
(straight rigid backbone),
we generalize the variant of the biased chain growth algorithm,
the pruned-enriched Rosenbluth method, for simulating
polymers with complex architecture, from star
polymers to bottle-brush polymers, on the simple cubic lattice.
With the high statistics of
our Monte Carlo results, we check the theoretical predictions of
side chain behavior and radial monomer density profile.
For the comparison of the experimental data for bottle-brush polymers with
a flexible backbone and flexible side chains, based on the bond fluctuation
model we propose another fast
Monte Carlo algorithm combining the local moves, the pivot
move, and an adjustable simulation lattice box.
By monitoring the autocorrelation functions of gyration radii for the
side chains and for the backbone, we see that for fixed side
chain length there is no change in the behavior of these two
functions as the backbone length increases. 
Our extensive results
cover the range which is accessible for the comparison to experimental
data and for the checking of the theoretically predicted scaling laws.

\end{abstract}

%% keywords here, in the form: keyword \sep keyword

%% PACS codes here, in the form: \PACS code \sep code

%% MSC codes here, in the form: \MSC code \sep code
%% or \MSC[2008] code \sep code (2000 is the default)

\maketitle

%%
%% Start line numbering here if you want
%%
% \linenumbers

%% main text
\section{Introduction}
\label{Introduction}

The so-called ``Bottle-Brush" polymers consist of a long molecule serving
as a ``backbone" on which many side chains are densely grafted.
The conformational change of bottle-brush polymers is mainly caused by 
the following factors: backbone length, side chain length, 
grafting density, type of monomers (chemical compound), and
solvent quality which can be adjusted by changing the temperature,
pH value, etc.  In the previous Monte Carlo studies of bottle-brush polymers,
both coarse-grained models on 
lattice~\cite{Rouault1996, Rouault1998, Shiokawa1999} and on 
off-lattice~\cite{Saariaho1997, Elli2004, Yethiraj2006}, show that
it is difficult to obtain high accuracy results for simulating
large bottle-brush polymers with high grafting densities.

On a coarse-grained scale, the bottle-brush polymer with densely grafted side 
chains may resemble a flexible long sphero-cylinder~\cite{Hsu2010a, Hsu2010b}. 
The complicated structure of bottle-brush polymers is therefore described 
in terms of multi-length scales such as the contour length $L_{cc}$,
the end-to-end distance of the backbone, $R_{eb}$, 
and of the side chain, $R_{e}$, the cross sectional radius $R_{cs}$,
and also the persistence length $\ell_p$ which describes the intrinsic
stiffness of the backbone, i.e., within the distance $\ell_p$, the 
cylinder is approximately straight. 
With computer simulations, one can estimate not only all these length
scales but also those physical quantities measured by experiments
such as the structure factors $S(q)$ which describe the scattering
function from any part of the bottle-brush polymers, and the radial
monomer density profile $\rho(r)$. Therefore, it is necessary to develop
an efficient algorithm for a deeper understanding of the
complicated structures of bottle-brush polymers 
with larger size and higher grafting densities in order to control
their functions for the applications in industry.

In this article, we first explain the
models and the algorithms for simulating bottle-brush polymers
with a straight rigid backbone, and with a flexible backbone in
Sec.~\ref{Model}. For simplicity, here we only focus on the case that
the bottle-brush polymers are under good solvent conditions.
In Sec.~\ref{Results} we present our results
and explain the connections between these estimates obtained by computer 
simulations, the theoretical predictions, and the experimental data, 
respectively.
Finally we give some conclusions in Sec.~\ref{Conclusions}.

%% The Appendices part is started with the command \appendix;
%% appendix sections are then done as normal sections
%% \appendix

%% \section{}
%% \label{}

\section{Models and algorithms}
\label{Model}  
    For studying bottle-brush polymers under a good solvent condition,
we first consider that bottle-brush polymers
consist of a rigid backbone and flexible side chains.
A simple coarse-grained model on a simple cubic lattice is used,
where the backbone is simply a rigid rod and flexible side chains are 
described by self-avoiding walks (SAWs) so that 
no multi-occupation of monomers on the same site is allowed. 
We apply a biased chain growth algorithm with resampling which is a variant
of the pruned-enriched Rosenbluth method (PERM)~\cite{g97, Hsu2004, Hsu2007} 
for the simulations. For the comparison between experimental data and
Monte Carlo simulations of bottle-brush polymers,
we need to consider a more complicated case that the backbone is
also flexible. The bond fluctuation 
model~\cite{Carmesin1988, Deutsch1991, Paul1991, Binder1995}
is used, where the backbone and all side chains are described by
SAWs on a simple cubic lattice but with some constraints 
(see sec.~\ref{bondfluctuation}).
We propose an algorithm which combines the local 26 moves,
the pivot moves and an adjustable simulation lattice box (LPB)
for the simulations~\cite{Hsu2010a, Hsu2010b,HsuPRL} 

\begin{figure*}[t]
\begin{center}
\includegraphics[width=.80\textwidth,angle=0]{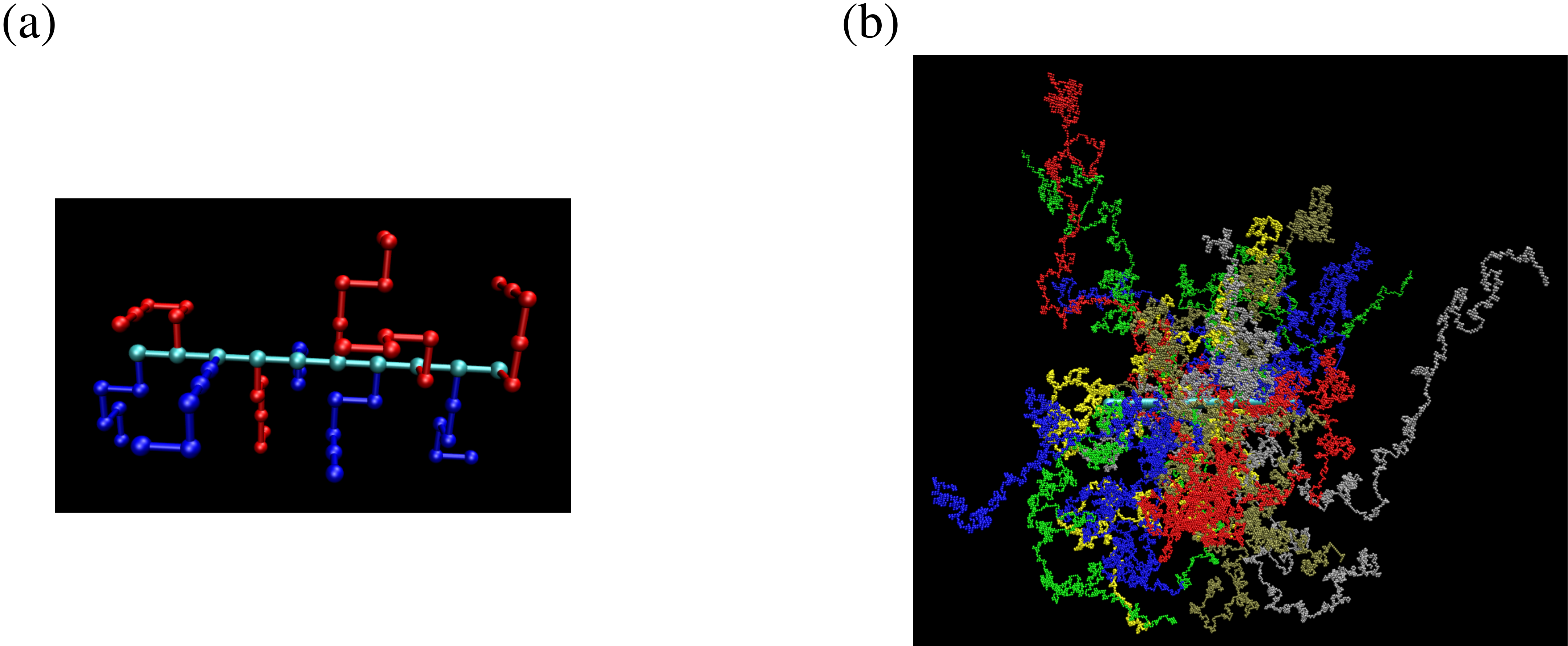}
\caption{(a) Schematic drawing of the geometric arrangement for
the simple coarse-grained model.
(b) A snapshot of a bottle-brush polymer with $N_b=128$,
$N=2000$, and $\sigma=1/4$ on the simple cubic lattice.
Note that different colors are used in order to distinguish between
different side chains, and the periodic boundary condition is undone
for the sake of better visualization.}
\label{fig-conf-PERM}
\end{center}
\end{figure*}

\subsection{Simple coarse-grained lattice model with PERM}

In the simple coarse-grained model, the backbone is fixed on the simple 
cubic lattice in the direction along the $z$-axis. 
$N_b$ monomers of the backbone are located on the lattice sites as 
shown in Fig.~\ref{fig-conf-PERM}. $n_c$ side chains consisting of $N$ monomers
each are grafted to the backbone monomers with equal distance $1/\sigma$ 
between two successive grafting sites on the backbone, where 
$\sigma$ is the grafting density defined by $\sigma=n_c/N_b$. 
Since mainly we want to check the scaling laws for very long side chains,
the periodic boundary condition is introduced in the direction along
the backbone to avoid the end-effects associated with a finite backbone 
length.
Differently from the conventional MC method where an initial configuration
is set up as a starter for the simulation, the conformation of
a bottle-brush polymer is built by growing all side chains simultaneously
with PERM. The partition sum of bottle-brush polymers of $n_c$ side chains 
of length $N$ each,
\begin{equation}
   Z_{Nn_c} = \sum_{walks} 1  \;,
\end{equation}
is therefore the total number of all possible configurations of
$n_c$ interactive self-avoiding random walks of steps $N$.
Only the excluded volume effect is considered here.

It is straight forward to apply the similar method for growing a star 
polymer~\cite{Hsu2004}
to the simulations for growing a bottle-brush polymer~\cite{Hsu2007}. 
Only in the latter case, side chains can either be attached to the same 
site or the different sites on the backbone depending on the grafting densities.
In the process of growing a bottle-brush polymer, one has to be aware that 
both the interactions between monomers in the same side chain, and
the interactions between monomers on different side chains have
to be taken into account. If one side chain is grown entirely before the
next side chain is started, it will lead to a completely ``wrong"
direction of generating the configurations of a bottle-brush.
Therefore, one has to use the strategy that all side chains are grown
simultaneously. Namely, a monomer is added to each side chain step
by step until all side chains having the same length, then that the
next round of monomers is added.
After we labelled all monomers by numbers, it goes back to the
problem of growing a linear chains from the first monomer to the 
$(n_cN)${\it th} monomer ~\cite{g97}. Using PERM, the configurations
of bottle-brush polymers are built by adding one monomer at each step,
and each configuration carries its own weight.
A wide range of probability distributions can be used for selecting
one of the nearest neighbor free sites of each side chain end at the next step,
but the efficiency of the algorithm depends on the choice of the distribution.
For the current problem, the bias of growing side chains is used
by giving higher probabilities in the direction where there are more free 
next neighbor sites and in the outward directions perpendicular to the 
backbone, where the second part of bias decreases with the length of side chains
and increases with the grafting density.
The total weight $W_{m}$ $(m=nn_c)$ for a bottle-brush polymer 
of all side chains having the length $n$
with an unbiased sampling is determined recursively by
$W_{m}=\prod_{k=1}^{m} w_k=W_{m-1}w_{m}$. As the weight $W_{m}$
is gained at the $m{\it th}$-step with a probability $p_{m}$, one has
to use $w_{m}/p_{m}$ instead of $w_{m}$. By taking the average
of all possible configurations, the partition sum
\begin{equation}
        \hat{Z}_{m} = \frac{1}{M_{m}} \sum_{\alpha=1}^{M_{m}} 
W_{m}(\alpha)
\label{eq-Z}
\end{equation}
can be estimated directly, 
where $M_{m}$ is the total number of configurations $\{\alpha\}$.
This is the main advantage of using PERM.
For any observable $A_{m}$, the mean value is therefore,
\begin{equation}
   \bar{A}_{m} = \frac{1}{M_{m}} \frac{\sum_{\alpha=1}^{M_{m}}
A_{m} (\alpha)W_{m}(\alpha)}{\hat{Z}_{m}}
\end{equation}.

\begin{figure*}[t]
\begin{center}
\includegraphics[width=.95\textwidth,angle=0]{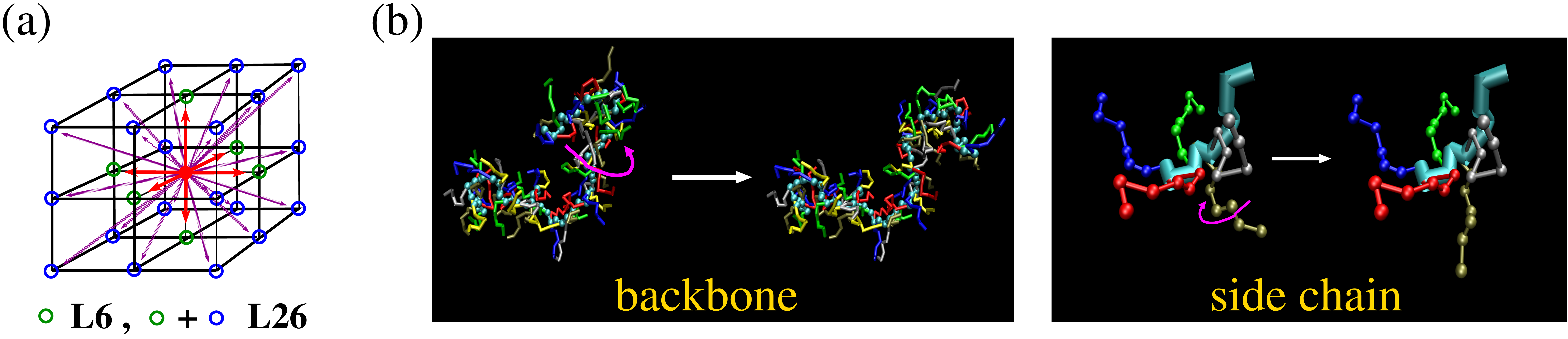}
\caption{(a) Schematic drawing of applying local 6 (L6) moves and
local 26 (L26) moves to a monomer on the site of a simple cubic lattice.
In the ``L6" moves, a monomer is tried to move to the nearest neighbor
sites in the six directions, while in the ``L26" moves, a monomer
is not only tried to move to the nearest neighbor sites but also to the
next nearest neighbor sites and the sites at the 8 corners which are
in $\sqrt{3}$ lattice spacings away from the chosen monomer.
(b) Two types of pivot moves applied to a randomly chosen monomer
on the backbone, and to a randomly chosen monomer on a randomly
chosen side chain.}
\label{fig-Lpb}
\end{center}
\end{figure*}

\begin{figure}[htb]
\begin{center}
\includegraphics[width=.35\textwidth,angle=0]{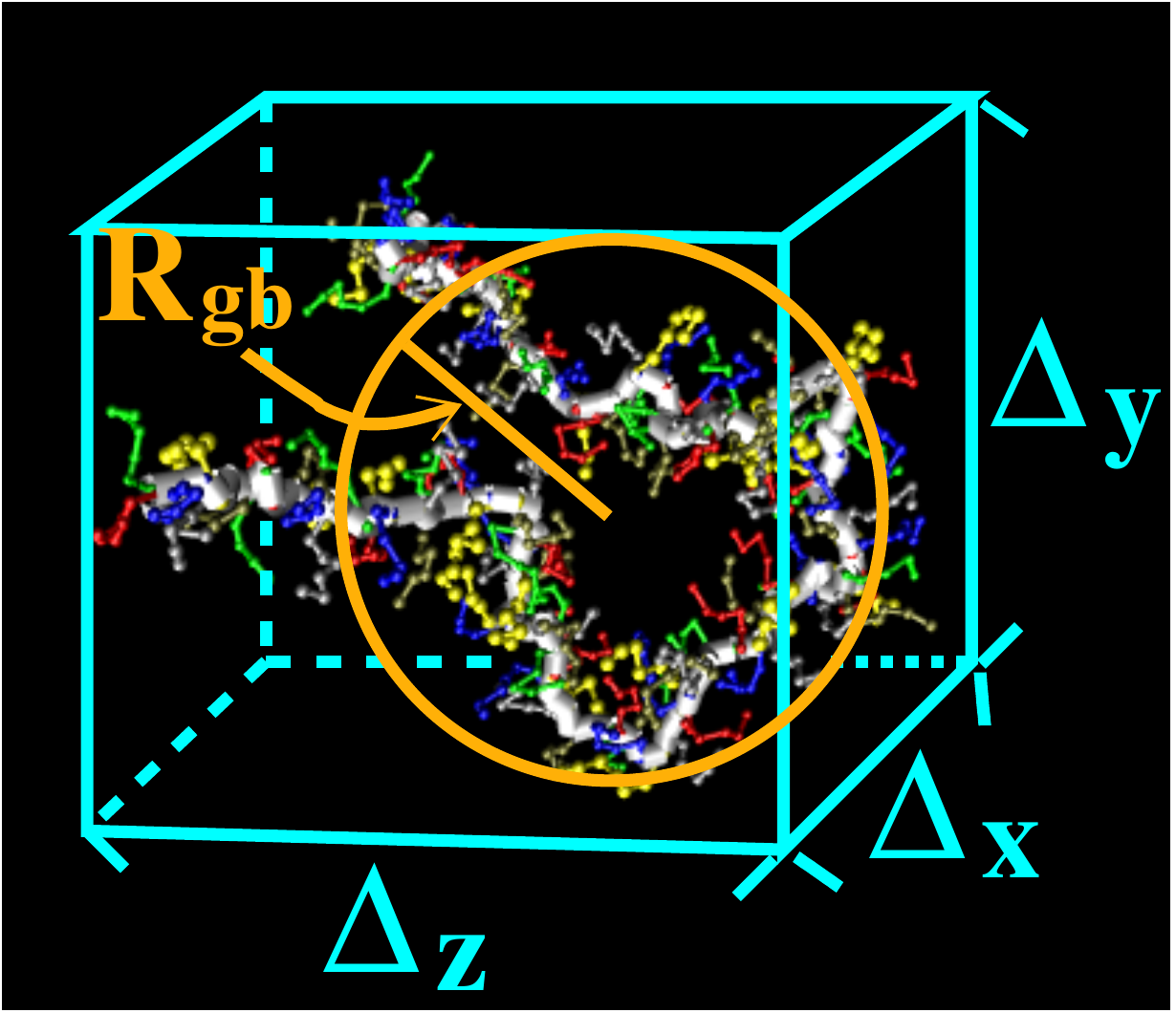}
\caption{The two observables, the radius of gyration of backbone
monomers, $R_{gb}$, and the space occupation of bottle-brush polymers,
$(\Delta_x, \Delta_y, \Delta_z)$, are indicated in the schematic drawing
by taking a snapshot of a bottle-brush polymer with $N_b=131$, $N=6$,
and $\sigma=1$.}
\label{fig-delta}
\end{center}
\end{figure}

In order to suppress the huge fluctuation of the probability distribution
and enrich those configurations with high weight the population control
is made in the way of pruning low weight configurations and cloning those
configurations with high weight.
Two thresholds $W_m^+$ and $W_m^-$ are introduced,
\begin{equation}
     W_m^+=C_+ \hat{Z}_m \, , \qquad W_m^-=C_- \hat{Z}_m
\end{equation}
where $\hat{Z}_m$ is the current estimate of the partition 
sum \{Eq.~(\ref{eq-Z})\}, and
$C_+$ and $C_-$ are constants of order unity. The optimal ratio between
$C_+$ and $C_-$ is found to be $C_+/C_- \sim 10$ in general.
For our simulations, we use $W_m^+=\infty$ and $W_m^-=0$ for the
first configuration hitting all side chains of length $n$.
For the following configurations, we use $W_m^+=C\hat{Z}_m(c_m/c_0)$ and
$W_m^-=0.15W_m^+$, here $C=3.0$, and $c_m$ is the total number of 
configurations of all side chains having length $n$. 
If the current weight $W_m(\alpha)> W_m^+$ for the configuration $\alpha$,
one produces two identical copies of this configuration, replaces their weight
$W_m(\alpha)$ by $W_m(\alpha)/2$. If $W_m(\alpha)<W_m^-$, 
one calls a random number $r$ where
$r \in [0,1]$. If $r \ge 1/2$, the configuration is kept but the weight 
is replaced by $2 W_m(\alpha)$, while the configuration is killed if $r<1/2$. 
Otherwise, the configuration is kept with the weight $W_m(\alpha)$.

A typical configuration of bottle-brush polymers under a good
solvent condition generated by PERM is shown in Fig.~\ref{fig-conf-PERM}(b).
It consists of $N_b=128$ backbone monomers, $N=2000$ side chain monomers
in each side chain, and the grafting density is $1/4$.
The total number of monomers is $N_{\rm tot}=N_b+N \sigma N_b=64128$.
So far, it is the largest bottle-brush polymers in the equilibrium state,
generated by MC simulations~\cite{Hsu2007}.

\subsection{Bond fluctuation model with LPB}
\label{bondfluctuation}

For studying bottle-brush polymers with a flexible backbone and flexible
side chains, we generalize the bond fluctuation model for a linear
polymer chain to that for a bottle-brush polymer. In the standard bond 
fluctuation model~\cite{Carmesin1988, Deutsch1991, Paul1991, Binder1995},
a flexible polymer chain is described by a SAW on a 
simple cubic lattice with bond constraints.
Each effective monomer blocks all 8 corners of an elementary cube
of the lattice from further occupation. Two successive monomers along a
chain are connected by a bond vector chosen from the set
$\{(\pm 2,0,0)$, $(\pm 2, \pm 1, 0)$, $(\pm 2, \pm 1, \pm 1)$, 
$(\pm 2, \pm 2, \pm 1)$, $(\pm 3,0,0)$, $(\pm 3,\pm 1,0)\}$,  
including also all permutations.
The geometry of a bottle-brush polymer with $N_b$ backbone monomers,
and with $n_c$ side chains of length $N$ is arranged in
the way that side chains are added to the backbone chain
at regular spacing $1/\sigma=N_b/n_c$, and two additional monomers are
added to each chain end of the backbone. Thus, the total number of
monomers is $N_{\rm tot}=\left[(n_c-1)/\sigma+1\right]+2$.
For our simulations, one of the simplest ways to set up the initial 
configuration is to assume that the backbone and side chains all have 
rod-like structures. 
Placing the backbone along the $z$-direction and fixing the bond length 
between two successive backbone monomers to be $3$, and randomly choosing the 
bond vector of each side chain from one of the allowed bond vectors 
including all permutation in the xy-plane but keeping the bond vectors 
within each side chain fixed, the required condition of bond constraints 
is satisfied and no further check is needed.
In our algorithm, instead of trying to move a chosen monomer to 
the nearest neighbor sites named by the local 6 (``L6") moves for the standard
bond fluctuation model, we use the local 26 (``L26") moves~\cite{Wittmer}
where it is tried to move to the $26$ neighbor sites as
shown in Fig.~\ref{fig-Lpb}(a). The local move is only accepted if 
the selected site is empty and the bond length constraints
are satisfied. In addition, two types of pivot moves are attempted.
One is that a monomer is chosen randomly on the backbone and
the short part of the bottle-brush polymer is transformed by randomly
applying one of the 48 symmetry operations (no change; rotations by
$90^o$ and $180^o$; reflections and inversions).
The other is that a monomer is chosen randomly from all side chain
monomers, and the part of side chain from the selected monomer to the
free end of the side chain is transformed by one of the 48 symmetry operations.

We first make some test runs for a small bottle-brush polymers
with $N_b=32$, $N=6$, $12$, $24$, and $48$, and $\sigma=1$ in order to 
compare the efficiency between the ``L26" moves and the
``L26" moves + pivot moves.
For any observable $A$, the performance of the algorithm is determined by 
the autocorrelation function $c(A,t)$,
\begin{equation}
   c(A,t)=\frac{\langle A(t_0)A(t_0+t)\rangle - \langle A(t_0) \rangle
\langle A(t_0+t) \rangle }{\langle A(t_0)^2 \rangle 
- \langle A(t_0) \rangle^2} \;.
\end{equation}
Results of $c(A,t)$ for the mean square gyration radius of the backbone,
$A=R_{gb}^2$, and of the side chains, $A=R_{gc}^2$ (taking the average of
all side chains at the same MC step $t$) plotted against
the number of MC steps $t$ show that 
the ``L26" + pivot algorithm is two orders of magnitude
faster than the ``L26" algorithm for fixed side chain length $N$
(figures are not shown here).
In the ``L26" algorithm, one MC step consists of $N_{\rm tot}$ ``L26"
moves, i.e., each monomer is selected once for the local move.
In the ``L26" + pivot algorithm, one MC step consists of $N_{\rm tot}$
``L26" moves, $k_{b}$ times pivot moves of the backbone and $k_{c}$ times
pivot moves of side chains. $k_{b}$ is chosen such that the acceptance ratio 
is about 40\% or even larger, while $k_c$ is $n_c/4$.

\begin{figure}[htb]
\begin{center}
(a)\includegraphics[width=.31\textwidth,angle=270]{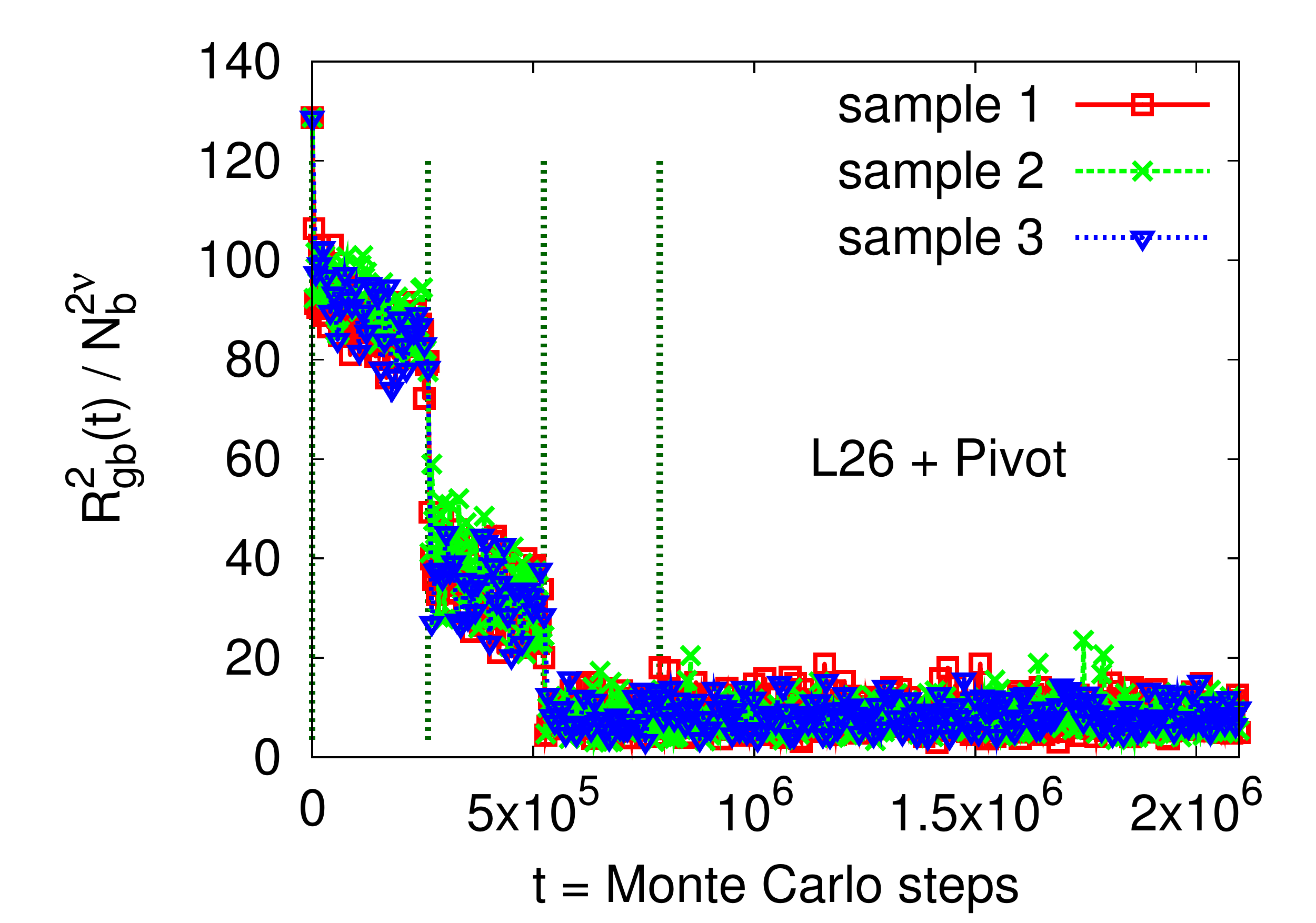}\hspace{1.8mm}
(b)\includegraphics[width=.31\textwidth,angle=270]{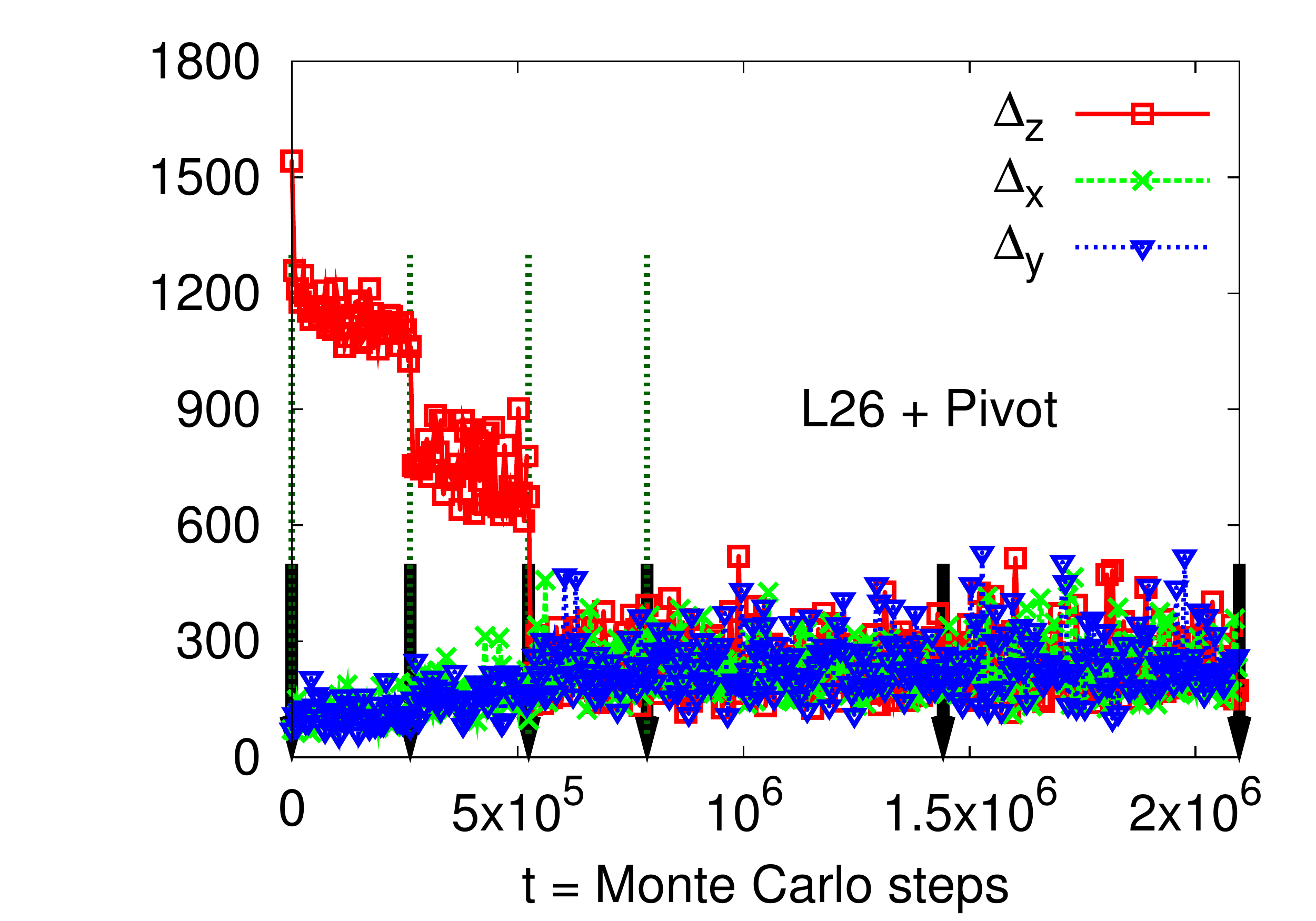}
\caption{Time series of the rescaled square gyration radii for the
backbone monomers, $R^2_{gb}(t)/N_b^{2\nu}$ (a) and
of the spacing occupation of the whole bottle-brush polymers in the Cartesian
coordinates, $(\Delta_x(t),\Delta_y(t),\Delta_z(t))$ (b).
Results shown here are for bottle-brush polymers with $N_b=515$ backbone
monomers, $N=12$ side chain monomers, and the grafting density $\sigma=1$.}
\label{fig-515-t}
\end{center}
\end{figure}

\begin{figure*}[t]
\begin{center}
\includegraphics[width=.80\textwidth,angle=0]{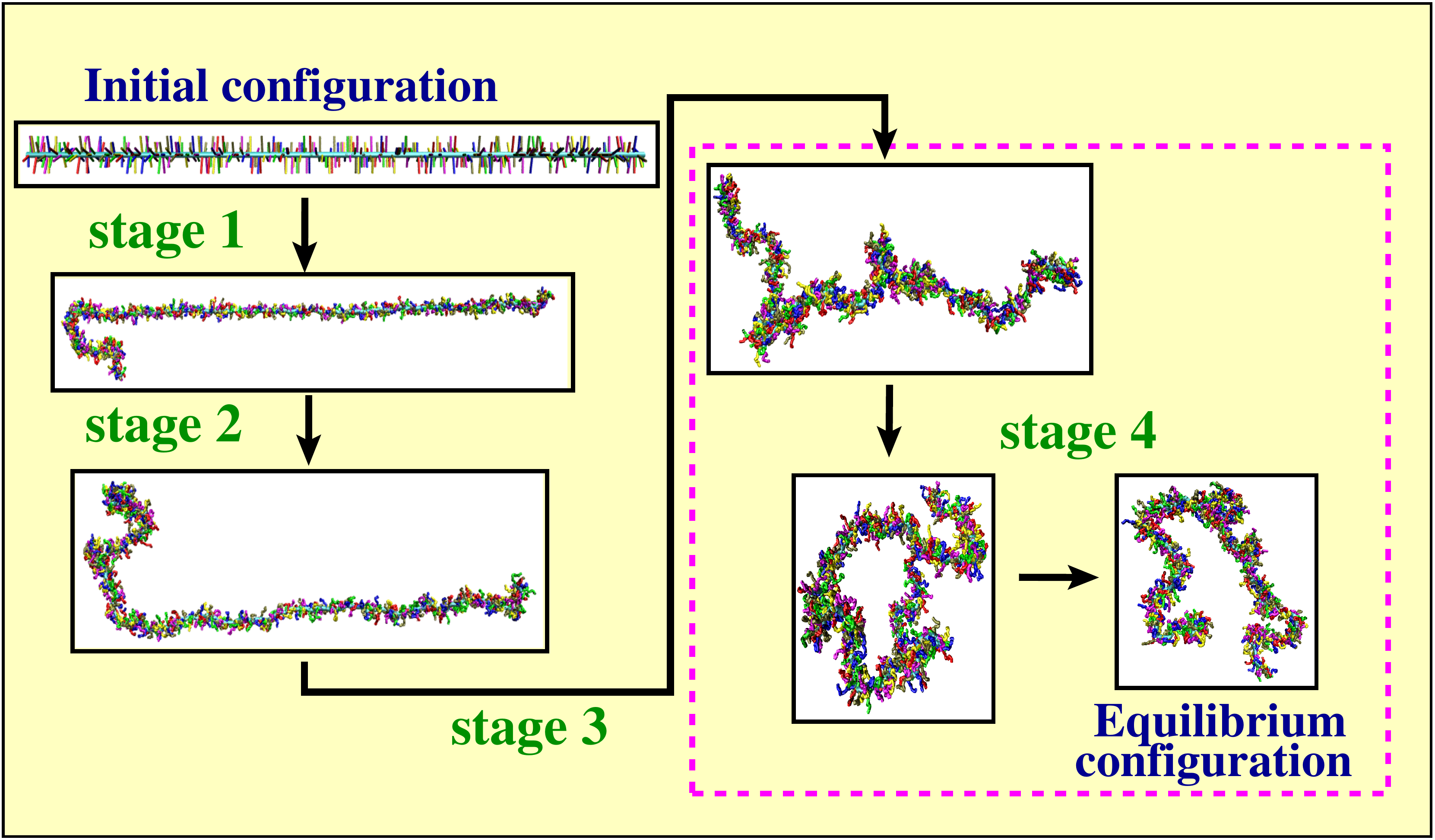}
\caption{Snapshots of bottle-brush polymers with $N_b=515$,
$N=12$, and $\sigma=1$ in the equilibrating process including
four stages.}
\label{fig-ie}
\end{center}
\end{figure*}

Two observables chosen for monitoring the equilibrating process
as shown in Fig.~\ref{fig-delta} are the radius of gyration of 
the backbone
monomers, $R_{gb}(t)$, and the space occupation of the bottle-brush 
polymer in the Cartesian coordinates, 
$(\Delta_x(t), \Delta_y(t), \Delta_z(t))$, 
since the average conformations of bottle-brush
polymers in equilibrium must be isotropic.
For simulating small bottle-brush polymers, we can simply set 
the three orthogonal length scales in the Cartesian coordinates
having equal length, e.g. $L_x=L_y=L_z=3N_b$ in the equilibrating process.
As the size of bottle-brush polymer increases, we will meet the 
problem of setting the simulation lattice box in our simulations due to the
limitation of the computer memory. The maximum volume of the box is 
$V=L_xL_yL_z=2^{28}$ for those computers we can access.
The solution for it is to adjust the simulation lattice box during 
the equilibrating process and separate the process into 
several stages.
Let's take a bottle-brush polymer with $N_b=515$, $N=12$, and $\sigma=1$
as an example.
The equilibrating process is separated into four stages as follows,
\begin{description}
\item[stage 1:] $1 \le N_b^p \le 128$,
$L_z^{(1)}=1545$, $L_y^{(1)}=L_x^{(1)} = 415$, $t_f^{(1)}=262144$ MC steps
\item[stage 2:] $1 \le N_b^p \le 256$, $L_z^{(2)}=1201$,
$L_y^{(2)}=L_x^{(2)} = 473$, $t_f^{(2)}=262144$ MC steps
\item[stage 3:] $1 \le N_b^p \le 513$, $L_z^{(3)}=851$,
$L_y^{(3)}=L_x^{(3)} = 561$, $t_f^{(3)}=262144$ MC steps
\item[stage 4:] $1 \le N_b^p \le 513$, $L_z^{(4)}=L_y^{(4)}=L_x^{(4)} = 645$,
$t_f^{(4)}=1310720$ MC steps
\end{description}
Here $N_b^p$ is the pivot point selected from the backbone monomers.
One has to be aware that
the pivot points which can be selected for applying the
pivot moves are also limited due to the current set up of the 
simulation lattice box.
However, once the system is in equilibrium one has to allow all possible 
moves, i.e. $(1 \le N_b^p \le (N_b-2))$.
At every new stage $k$, the initial configuration is taken from the last
stage, and the size of the simulation lattice box is decided by the
final space occupation of the bottle-brush polymers at the last stage,
i.e. $L_z^{(k)} \ge \Delta_z(t)$, 
$L_y^{(k)}=L_x^{(k)} \approx \max (\Delta_y(t), \Delta_x(t))$, with
$t=\sum_{i=1}^k t_f^{(i-1)}$, and $L_z^{(k)}L_y^{(k)}L_x^{(k)}\le 2^{28}$.
Time series of $R_{gb}(t)$ and $(\Delta_x(t), \Delta_y(t), \Delta_z(t))$
are shown in Fig.~\ref{fig-515-t}. 
The four stages are separated by the vertical green curves.
Taking some snapshots of the conformations of the bottle-brush
polymers at the Monte Carlo steps indicated by the arrows in 
Fig.~\ref{fig-515-t}(b) one can see how the conformations of 
bottle-brush polymers change during the equilibrating process as
shown in Fig.~\ref{fig-ie}.
At the beginning, backbone and side chains are in rod-like
structures. At the end of the first stage, only a small part of the backbone is
flexible. As more backbone monomers are relaxed, we see that 
the backbone and the side chains become more and
more flexible step by step. Finally an equilibrium state is reached.
It takes about $1.25$ hours CPU time on an Intel 2.8 GHZ PC for 
such a bottle-brush polymer to reach the equilibrium state by
choosing $k_b=40$ and $k_c=128$.

\begin{figure}[htb]
\begin{center}
(a)\includegraphics[width=.31\textwidth,angle=270]{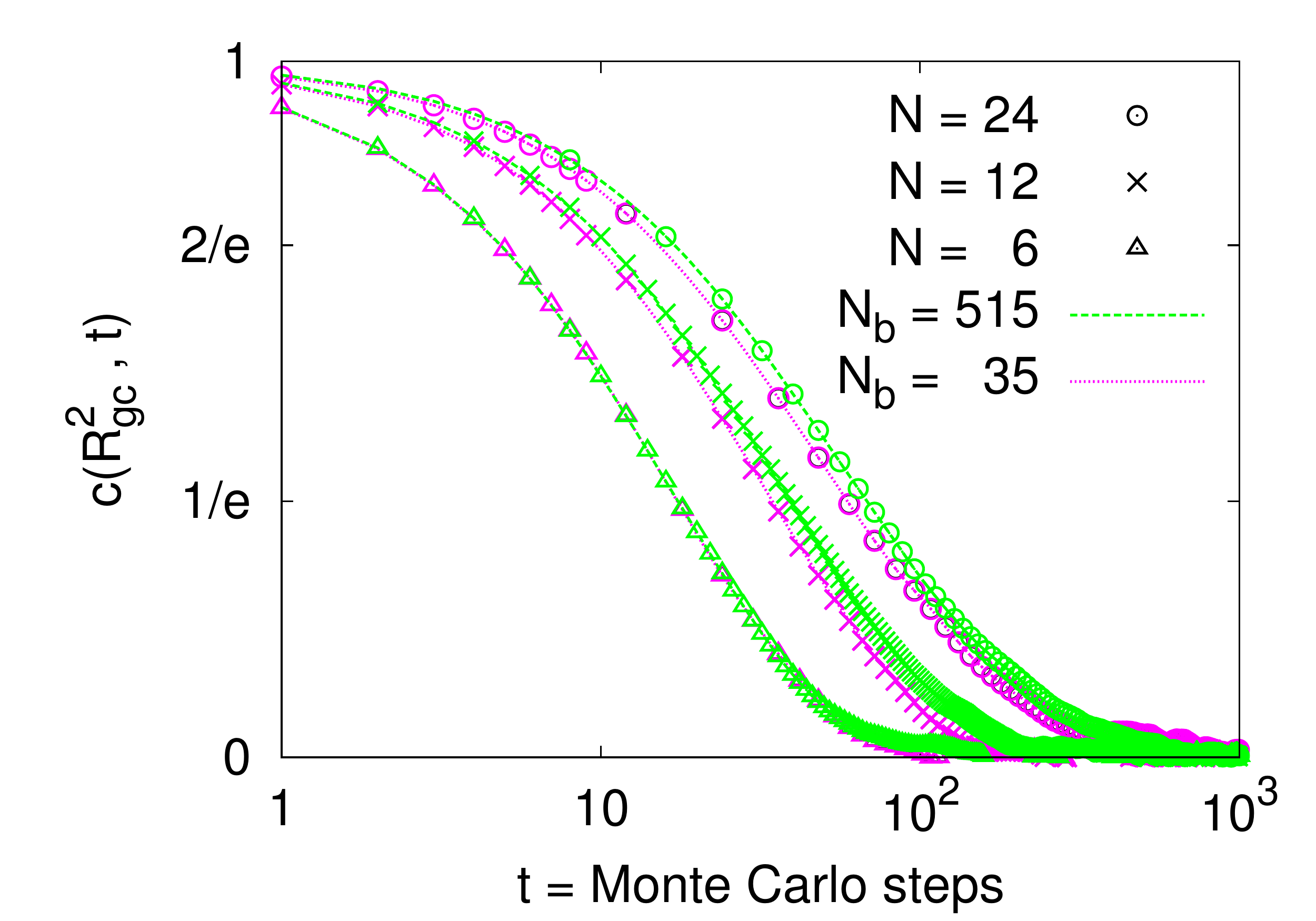}\hspace{1.8mm}
(b)\includegraphics[width=.31\textwidth,angle=270]{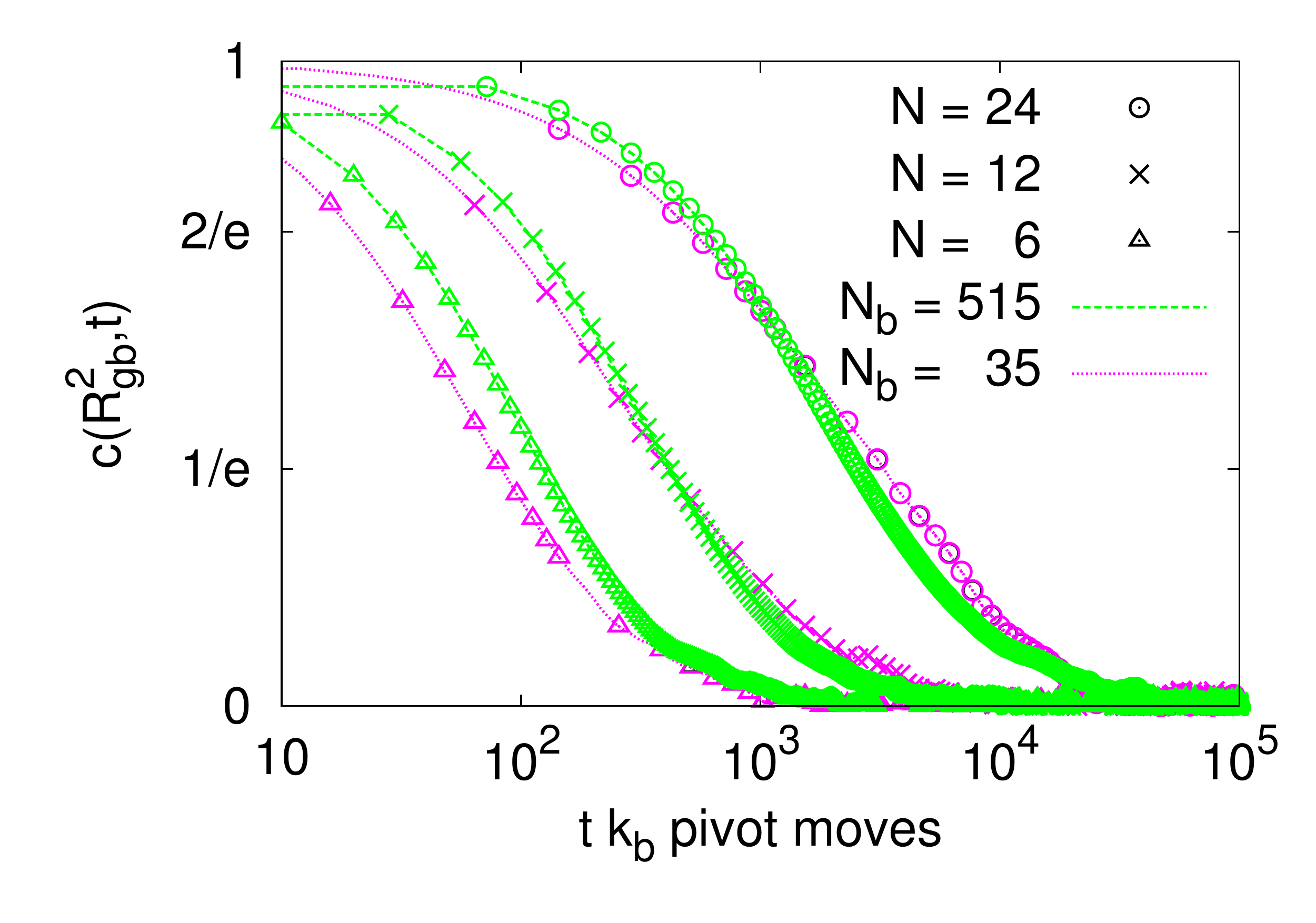}
\caption{Autocorrelation functions of the mean square gyration radii for
the side chains $c(R^2_{gc}, t)$ (taking the average of all side chains
at each $t$) plotted against the number of Monte Carlo steps $t$ (a),
and for the backbone $c(R^2_{gb}, t)$ plotted against the number of
pivot moves applied to the backbone, $tk_b$ (b).
Data obtained by the algorithm LPB for bottle-brush polymers with
$N_b = 515$, and $N_b=35$ are shown by dashed and dotted curves, respectively.}
\label{fig-c-515}
\end{center}
\end{figure}

It is more time consuming when pivot moves are applied 
to the simulations for larger bottle-brush polymers.
In order to know how much the efficiency is slowing down as
the backbone length increases, we compare the autocorrelation functions
$c(R^2_{gc},t)$ and $c(R^2_{gb},t)$ for bottle-brush polymers with
$N_b=35$ and $N_b=515$ for three different side chain lengths 
$N=6$, $12$, and $24$.  
For both cases in Fig.~\ref{fig-c-515}(a), we see that 
the decay of the autocorrelation function for the side chain 
structure occurs on the same time scale. It is also true for the
backbone as shown in Fig.~\ref{fig-c-515}(b) that the autocorrelation
functions are plotted against the number of pivot moves $tk_b$.
Clearly, the structural relaxation time is longer as the side chain length
$N$ increases~\cite{Hsu2011}.

\section{Results}
\label{Results}

\begin{figure}[t]
\begin{center}
(a)\includegraphics[width=.31\textwidth,angle=270]{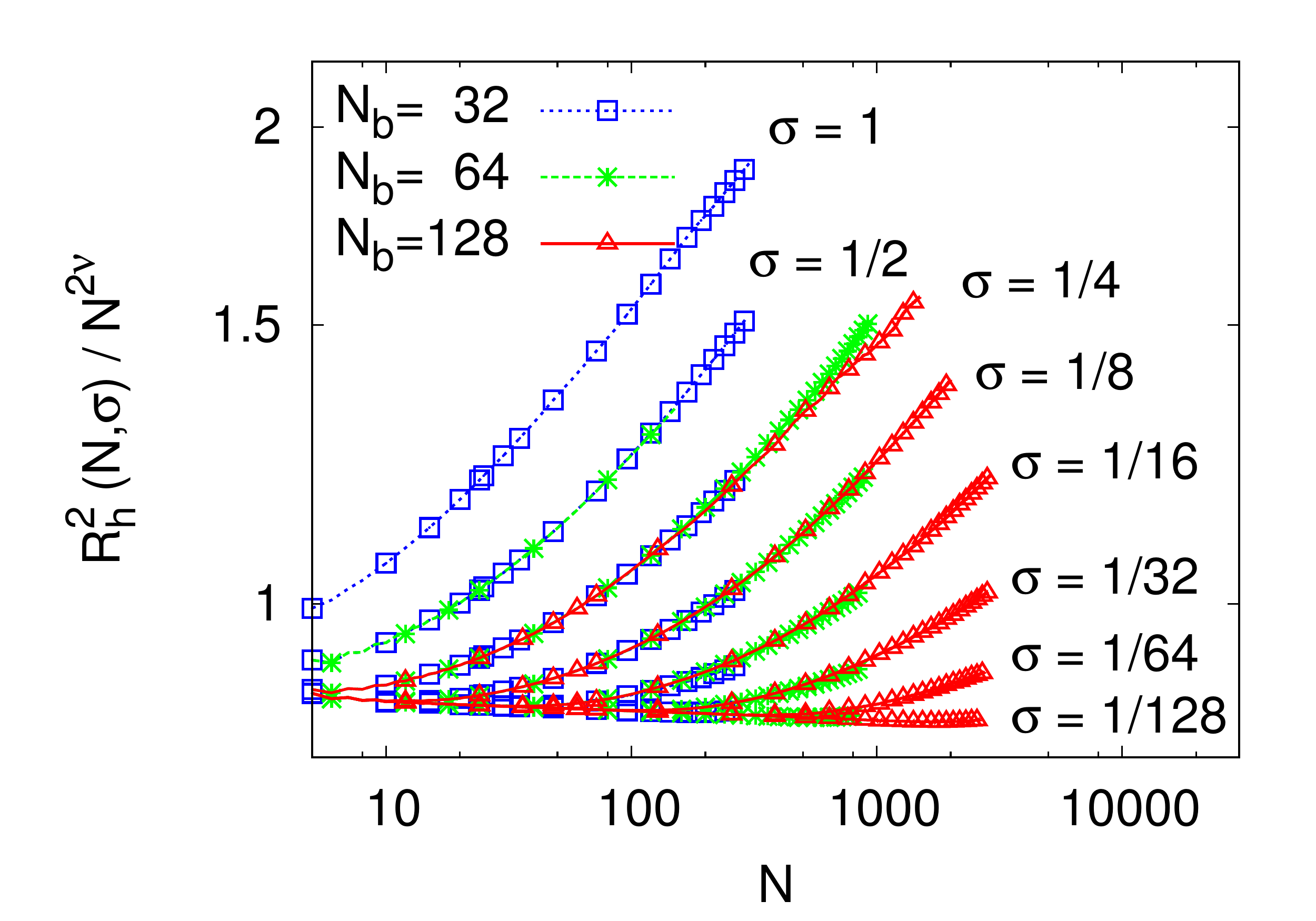}\hspace{1.8mm}
(b)\includegraphics[width=.31\textwidth,angle=270]{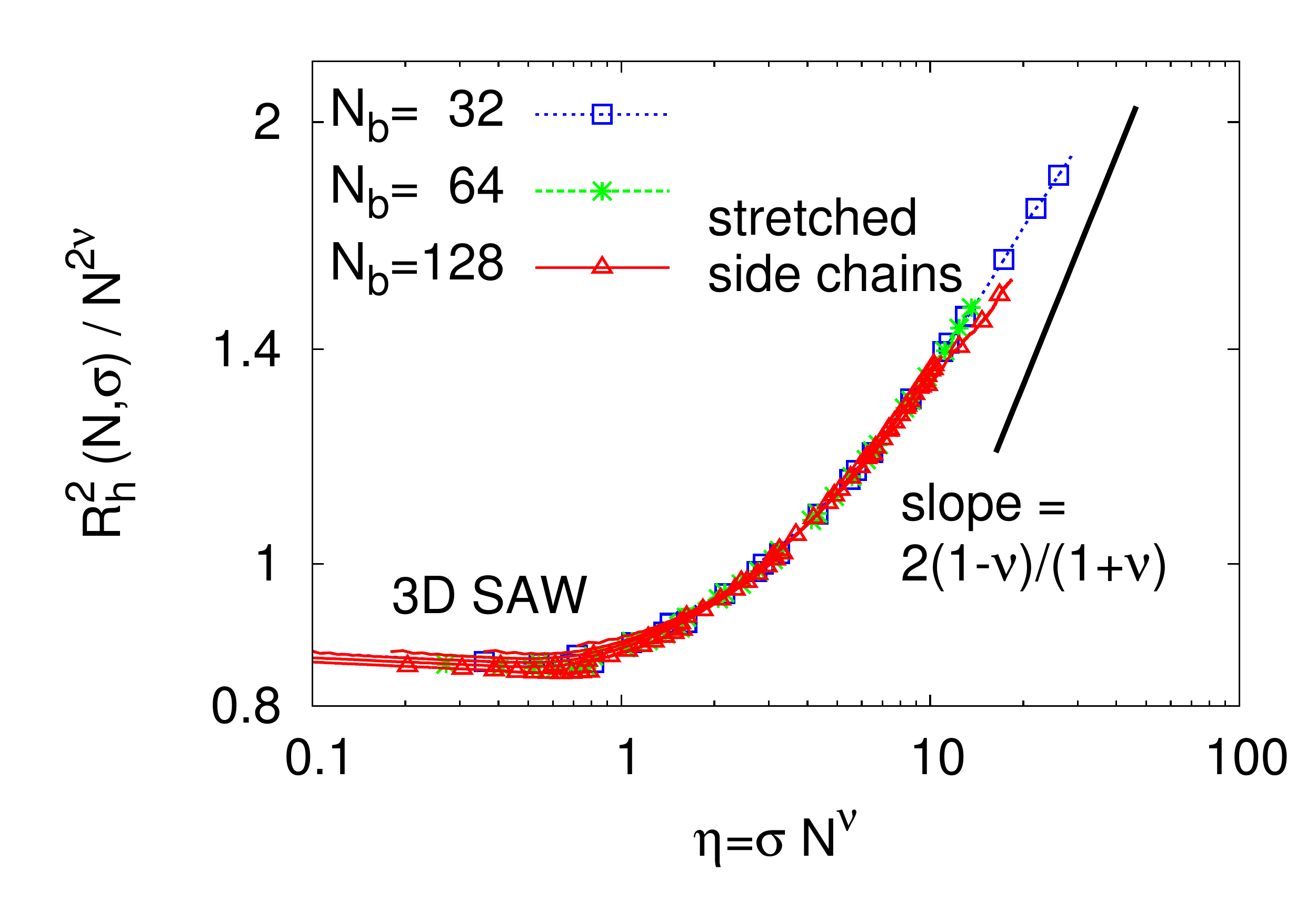}
\caption{Log-log plot of rescaled mean square height $R_h^2(N,\sigma)/N^{2 \nu}$
versus $N$ (a) and $\eta=\sigma N^\nu$ (b) with $\nu=0.588$.
Results are obtained for three choices of $N_b$ and several choices of the
grafting density $\sigma$ as indicated. Those unphysical data
($R_h>0.5N_b$) due to the
artifact of using periodic boundary condition are removed.
The slope of the straight line corresponds to the scaling estimate from
Eq.~(\ref{eq-height}).}
\label{fig-Reexy}
\end{center}
\end{figure}

According to the cylindrical geometry of bottle-brush polymers with
a rigid backbone, we extend the Daoud-Cotton blob picture for a star polymer
to that for a bottle-brush polymer. The space is partitioned into blobs of
non-uniform size and shape. The blobs are not spheres but rather ellipsoids.
Based on this theory, the scaling law for side chains in the 
radial direction (height of the bottle-brush) is given by,
\begin{equation}
    R_h(N,\sigma) \propto  \sigma^{(1-\nu)/(1+\nu)}N^{2\nu/(1+\nu)} \,, \qquad
{\rm for} \, \sigma \rightarrow \infty
\label{eq-height}
\end{equation}
where $\nu$ is the Flory exponent for 3D SAWs ($\nu \approx 0.588$).
With the first part of 
simulations by using PERM, results of the mean square height of 
bottle-brush polymers for three choices of backbone length $N_b=32$,
$64$, and $128$, and several choices of the grafting density $\sigma$ 
from $1/128$
to $1$ are shown in Fig.~\ref{fig-Reexy}(a). 
We see that those curves of the same grafting density $\sigma$
coincide with each other. Increasing the grafting density $\sigma$
enhances the stretching of side chains.
Considering that in the mushroom
regime $(\sigma \rightarrow 0)$, the height of bottle-brush polymers should
behavior as 3D SAWs, i.e. $R_h(N,\sigma\rightarrow 0) \sim N^{\nu}$, one
can write down the scaling ansatz in the thermodynamic limit 
as $N\rightarrow \infty$~\cite{Hsu2007},
\begin{equation}
    R^2_{h}(N,\sigma)=N^{2\nu}\tilde{R}^2(\eta)\, , \qquad \eta=\sigma N^{\nu}
\end{equation}
with
\begin{equation}
  \tilde{R}^2(\eta)=
\left\{\begin{array}{lll}
1 \, &, & \eta \rightarrow 0 \\
\eta^{2(1-\nu)/(1+\nu)}\, &, &
\eta \rightarrow \infty\\
\end{array}
\right .
\end{equation}
For checking this cross-over scaling ansatz, we plot the
same data as shown in Fig.~\ref{fig-Reexy}(a), but rescale the x-axis 
from $N$ to $\eta$.
We see the nice data collapse. As $\eta$ increases, a
cross-over from a 3D SAWs to a stretched side chain regime is indeed seen, but
only rather weak stretching of side chains is realized, which is different
from the scaling prediction \{Eq.~(\ref{eq-height})\}. In this log-log plot, the
straight line gives the asymptotic behavior of the scaling prediction
for very large $\eta$.
However, this is the first time we can see the cross-over behavior
by computer simulations. This cross-over regime is far from reachable by
experiments.
On the other hand, it requires a lot of effort to reach the regime where the
theoretical prediction would apply, either the grafting density $\sigma$ has
to be much higher or the side chain length $N$ has to be much longer.

\begin{figure}[htb]
\begin{center}
(a)\includegraphics[width=.31\textwidth,angle=270]{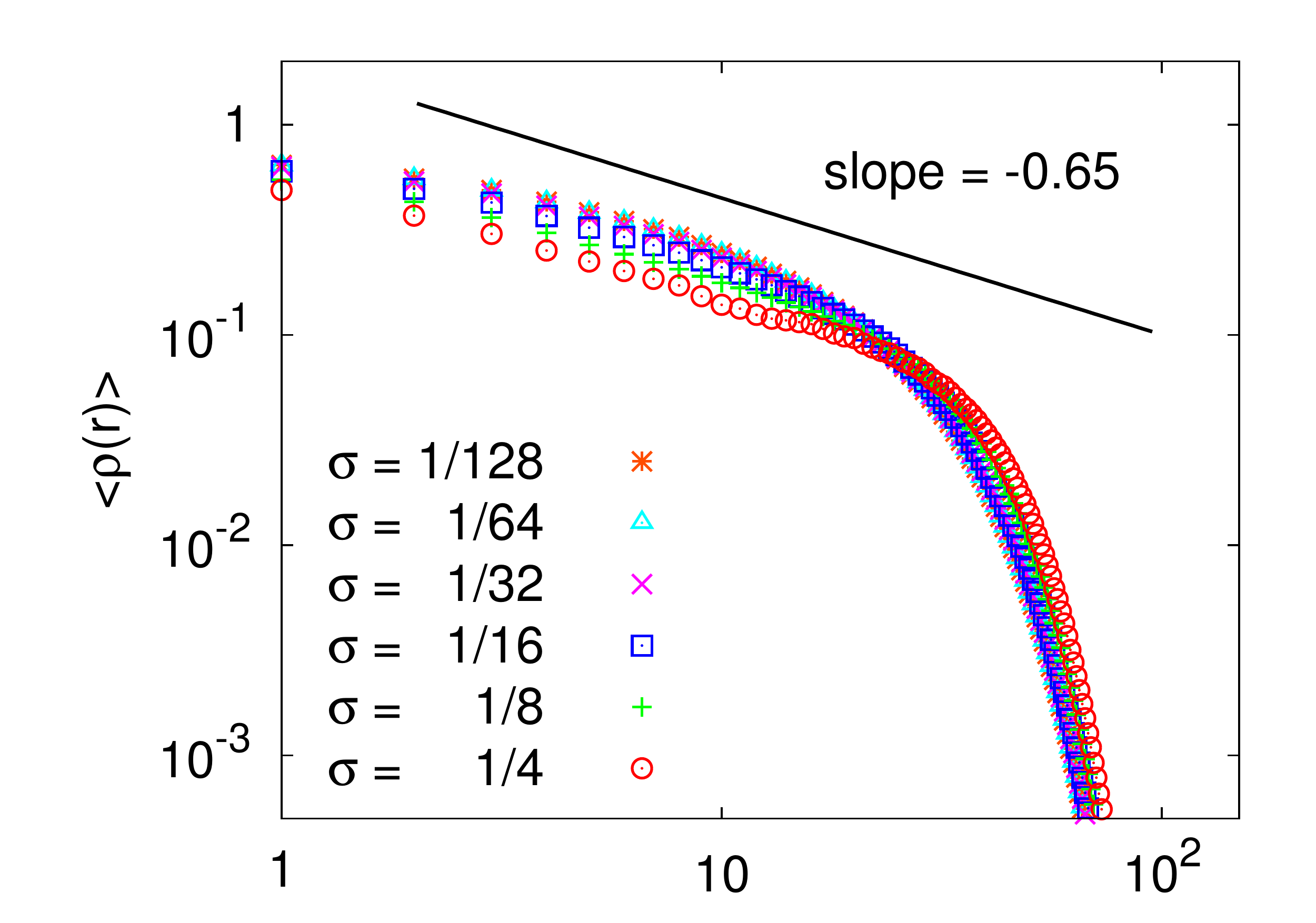}\hspace{1.8mm}
(b)\includegraphics[width=.31\textwidth,angle=270]{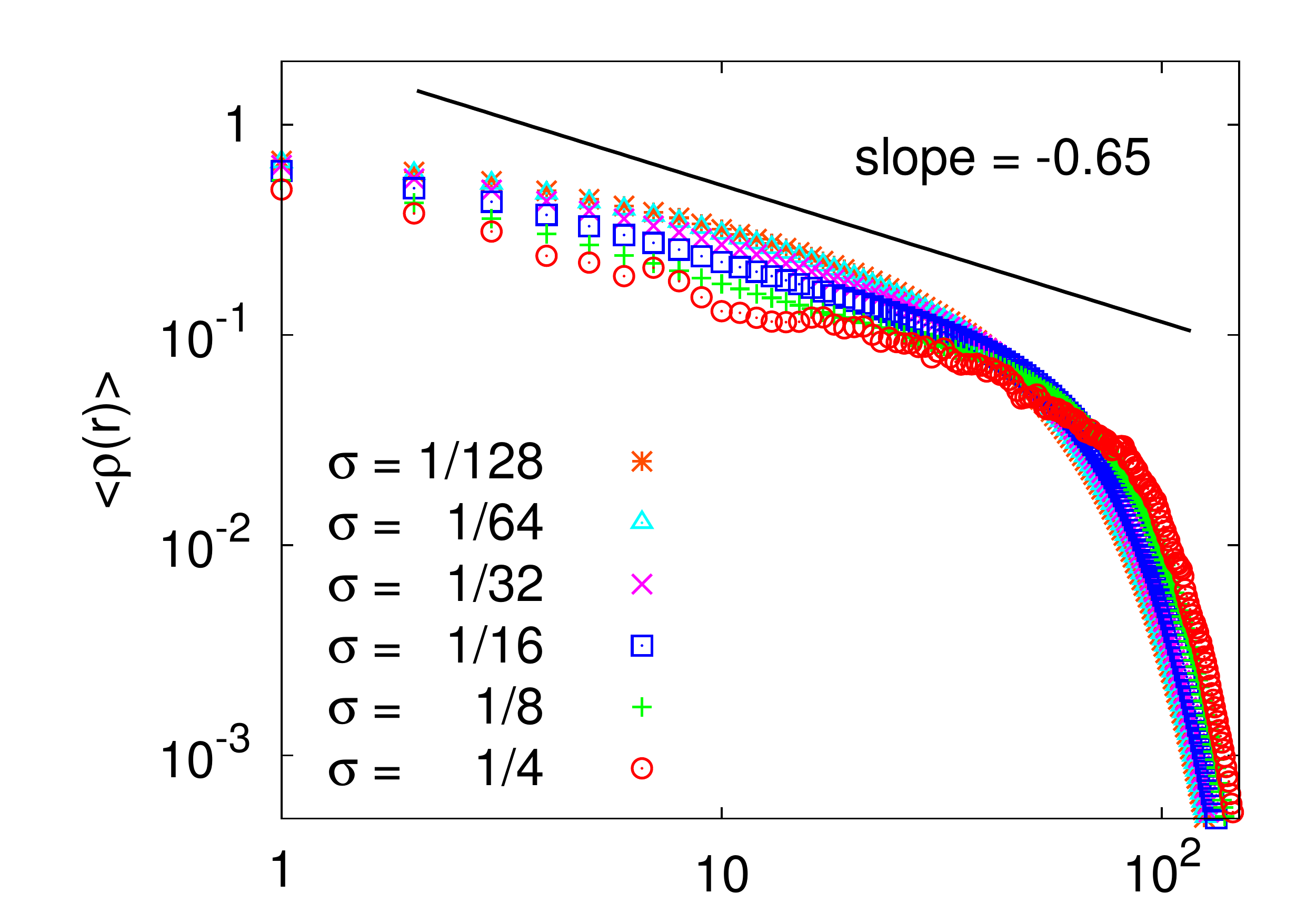}
\caption{Radial distribution function $\rho(r)$ plotted against $r$,
for $N=500$ (a) and $N=1500$ (b).}
\label{fig-rho}
\end{center}
\end{figure}

\begin{figure*}[htb]
\begin{center}
\includegraphics[width=.90\textwidth,angle=0]{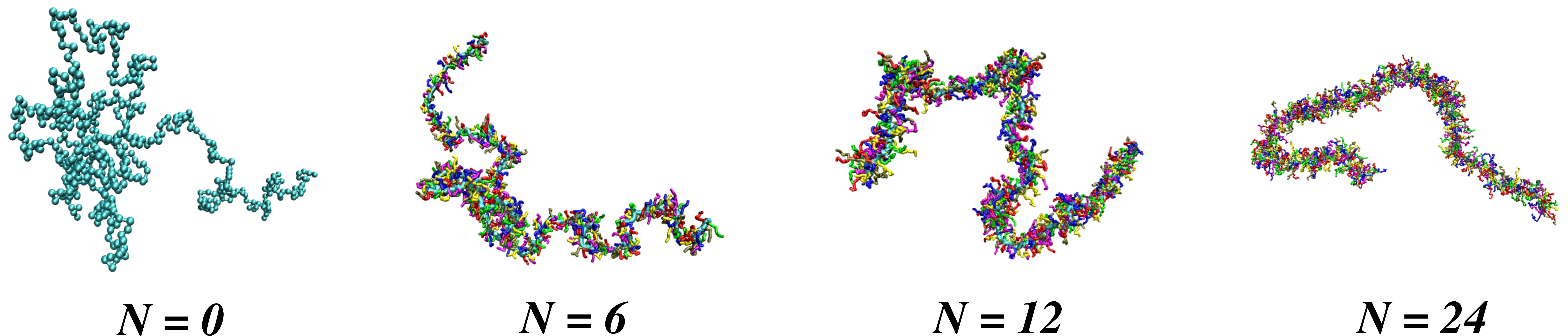}
\caption{Snapshots of bottle-brush polymers with $N_b=515$ backbone
monomers, $N$ side chain monomers, and with the grafting density
$\sigma=1$. As $N$ increases from $0$ (linear polymer chain) to
$24$, the backbone becomes stiffer. Quantitatively, the local
intrinsic stiffness of the backbone is described by the persistence
length $\ell_p$ as shown in Fig.~\ref{fig-lp}(a).}
\label{fig-515-snap}
\end{center}
\end{figure*}

The same situation is also observed as we check the scaling prediction for the
radial distribution function~\cite{Hsu2007}
\begin{equation}
    \rho(r) \propto (r/\sigma)^\delta \,, \qquad
 \delta=\frac{1-3\nu}{2\nu} \approx -0.65
\end{equation}
Results are shown in Fig.~\ref{fig-rho} for bottle-brush polymers with
side chain length $N=500$ and $N=1500$. As we keep all the grafting
densities $\sigma$ fixed but increase the side chain length $N$,
one can see only in a rather tiny regime, the data seem to follow the same
slope as predicted by the theory. Therefore, one might expect that finally the
radial distribution function would follow the predicted scaling law as
both $\sigma$ and $N$ are very large.

\begin{figure*}
\begin{center}
(a)\includegraphics[width=.31\textwidth,angle=270]{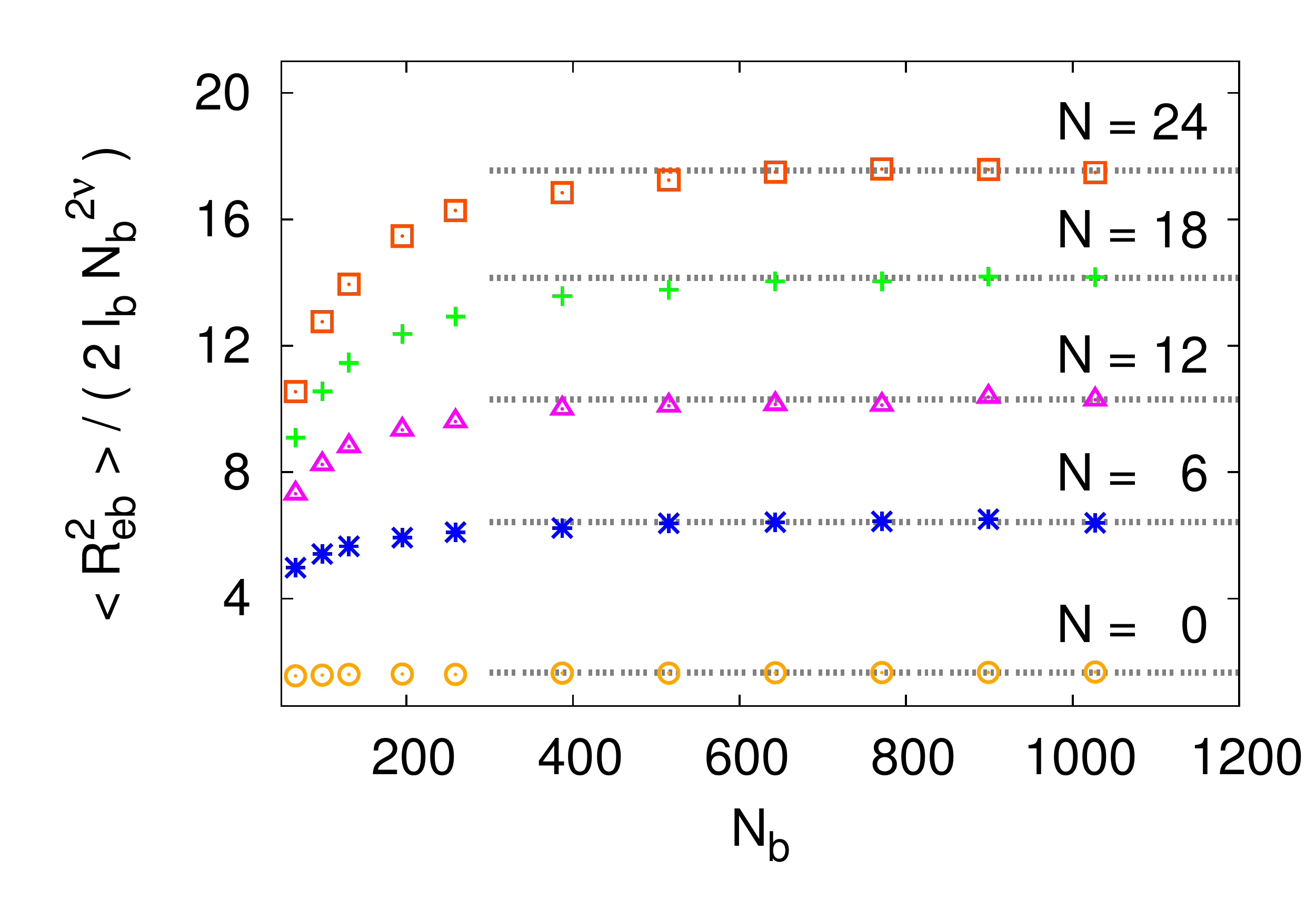}\hspace{1.8mm}
(b)\includegraphics[width=.31\textwidth,angle=270]{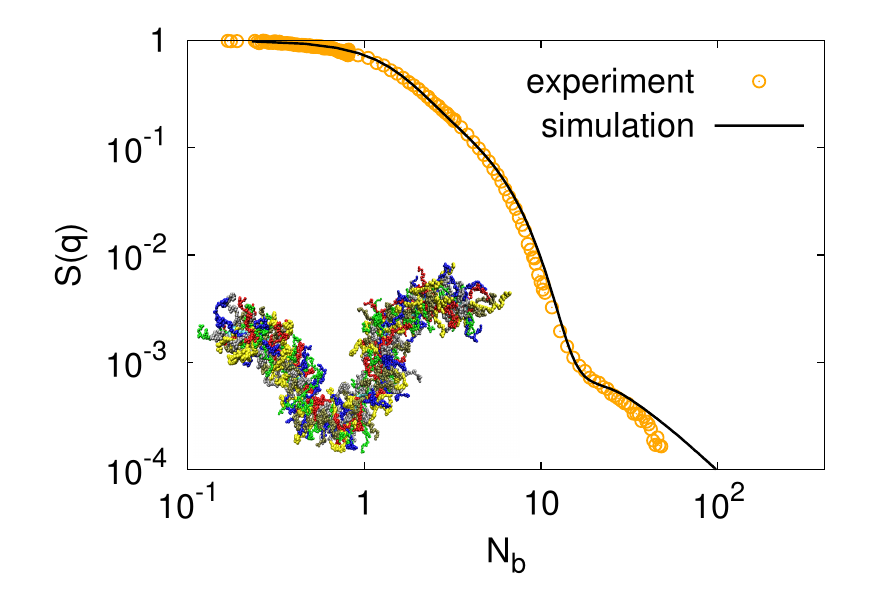}
\caption{(a) Rescaled mean square end-to-end distance of the bottle-brush
polymers, $\langle R_{eb}^2 \rangle /(2 \ell_b N_b^{2\nu})$, 
plotted against the number
of monomers on the backbone, $N_b$, for the grafting density $\sigma=1$,
and several choices of side chain length $N$.
The corresponding persistence length $\ell_p$ for fixed side chain lengths $N$
are given by the horizontal curves.
(b) Structure factors $S(q)$ plotted against the wave factor $q$.
Simulation results are obtained for the bottle-brush polymer with $N_b=259$,
$N=48$, and $\sigma=1$ by LPB. Experimental data for the sample {\bf B2}
with $N_b^{({\rm exp})}=400$ and $N^{({\rm exp})}=62$
are quoted from Ref.~\cite{Rath}. A snapshot of the bottle-brush polymer
generated by LPB is also shown in (b).}
\label{fig-lp}
\end{center}
\end{figure*}

For the second part of the simulations, let's first look at the snapshots of 
bottle-brush polymers under a good solvent condition,
which contain $N_b=515$ backbone monomers, $N=0$ (linear polymer), 
$6$, $12$, and $24$ monomers on each side chain, and the grafting 
density $\sigma=1$.
As $N$ increases, one sees that the corresponding
conformations of bottle-brush polymers are rather different.
The backbone becomes stiffer as the side chain length increases.
This local intrinsic stiffness of the backbone
is quantitatively described by the persistence length $\ell_p$.
According to the scaling law of the mean square end-to-end distance of
the backbone~\cite{Hsu2010a,Hsu2010b},
\begin{equation}
    \langle R^2_{eb} \rangle =2 \ell_b \ell_p N_b^{2\nu} \, , \qquad {\rm as} \, 
N_b \rightarrow \infty
\end{equation}
where $\ell_b \approx 2.7$ is the average bond lengths for the bond
fluctuation model. Results of the rescaled mean square end-to-end distance
of the backbone for $N=0$, $6$, $12$, $18$, and $24$ and for various numbers
of backbone monomers $N_b$ are shown in Fig.~\ref{fig-lp}(a). 
The persistence length $\ell_p$ for fixed side chain length $N$ is 
determined by the plateau for large $N_b$ since finally those curves 
of $\langle R^2_{eb} \rangle/(2\ell_bN_b^{2\nu})$
all show a smooth cross-over from a rod-like chain
to a 3D SAW. $\ell_p$ increases as $N$ increases.

The most common quantity to describe the structure of
macromolecules is the structure factor $S(q)$ for the whole
bottle-brush polymer, which is estimated by taking the average 
of all independent
configurations obtained from MC simulations, i.e.
\begin{equation}
S(q)=\frac{1}{{\cal N}_{tot}} \sum_{i=1}^{{\cal N}_{tot}}
 \sum_{j=1}^{{\cal N}_{tot}} <c(\vec{r}_i)c(\vec{r}_j)>
\frac{\sin(q \mid \vec{r}_i - \vec{r}_j \mid)}
{q \mid \vec{r}_i - \vec{r}_j \mid}
\end{equation}
where $c(\vec{r}_i)=1$ if $\vec{r}_i$ is occupied, otherwise $c(\vec{r}_i)=0$. 
In experiments, $S(q)$ can be measured by using static light scattering,
small angle neutron scattering, and x-ray scattering, e.g. Ref.~\cite{Rath}.
By choosing the accessible size of bottle-brush polymers for experiments,
we have found the connections between our MC simulation results
and the experimental data~\cite{Hsu2010a}. 
One example is shown in Fig.~\ref{fig-lp}(b). 
As we normalize the structure factor $S(q)\rightarrow 1$ as 
$q\rightarrow 0$, and rescale the wave factor $q$ to $qR_g$ where
$R_g$ is the radius of gyration of the whole bottle-brush polymer, 
we see that the backbone length $N_b^{({\rm exp})}=400$
in the experiment corresponds to the backbone length
$N_b=259$ in the simulation, and side chain length $N^{({\rm exp})}=62$
in the experiment corresponds to the side chain length
$N=48$ in the simulation. The grafting density $\sigma \approx 1$ for
both cases. Immediately, we can translate
that $1$nm $\approx 3.79$ lattice spacings.

\section{Conclusions}
   In this paper, we study the bottle-brush polymers under good solvent
conditions by using two kinds of lattice models, a simple coarse-grained
model on the simple cubic lattice and the bond fluctuation model. 
Due to the complex characteristics of bottle-brush polymers, we
have proposed two algorithms, a variant of PERM and LPB depending on the 
interesting regime of length scales.
With our extensive MC simulations, we show that the stretching of 
side chains in the interior of the bottle-brush polymer is weaker
than the theoretical prediction. A convincing estimate of the
persistence length $\ell_p$ which describes the intrinsic stiffness
of bottle-brush polymers depending on the side chain length is given.
We also give a direct comparison of the structure factors between
our simulation results and the experimental data.
The newly developed algorithm LPB has also been employed successfully to study
the conformational change of bottle-brush polymers as they are adsorbed 
on a flat solid surface by varying the attractive interaction between the 
monomers and the surface~\cite{Hsu2010c}.
 
\label{Conclusions}

\noindent
{\bf Acknowledgement}

H.-P. H. received funding from the
Deutsche Forschungsgemeinschaft (DFG), grant No SFB 625/A3.
We are grateful for extensive grants of computer time at the JUROPA
under the project No HMZ03 and
SOFTCOMP computers at the J\"ulich Supercomputing Centre (JSC),
and PC clusters at ZDV, university of Mainz.
H.-P. H. thanks K. Binder and W. Paul for their kind support and 
collaboration. \\

\noindent 
{\bf References}
%%
%% Following citation commands can be used in the body text:
%% Usage of \cite is as follows:
%%   \cite{key}         ==>>  [#]
%%   \cite[chap. 2]{key} ==>> [#, chap. 2]
%%

%% References with BibTeX database:

\end{document}